\documentclass[conference]{IEEEtran}

\usepackage[utf8]{inputenc}
\usepackage[T1]{fontenc}

\usepackage{multirow}
\usepackage[inline]{enumitem}
\usepackage{xcolor}
\usepackage{booktabs}
\usepackage[pdfauthor={Maximilian Bachl, Joachim Fabini, Tanja Zseby}]{hyperref}
\usepackage{amsmath}
\usepackage{amssymb}
\usepackage{graphicx}
\usepackage[numbers]{natbib}
\usepackage{subfig}
\usepackage{tikz}
\usepackage{float}

\usepackage[nomain, toc, acronym]{glossaries}
\glsdisablehyper

\def\anonymize{no}
\def\tempYes{yes}
\def\tempNo{no}

\begin{document}

\title{Detecting Fair Queuing for Better Congestion Control}

\author{\IEEEauthorblockN{Maximilian Bachl, Joachim Fabini, Tanja Zseby}
\IEEEauthorblockA{Technische Universität Wien\\
firstname.lastname@tuwien.ac.at}}





\maketitle%

 \thispagestyle{plain}
 \pagestyle{plain}

\newacronym{cc}{CC}{Congestion Control}
\newacronym{cca}{CCA}{Congestion Control Algorithm}
\newacronym{aqm}{AQM}{Active Queue Management}
\newacronym{rtt}{RTT}{Round Trip Time}
\newacronym{fq}{FQ}{Fair Queuing}
\newacronym{bdp}{BDP}{Bandwidth Delay Product}
\newacronym{qoe}{QoE}{Quality of Experience}

\begin{abstract}
Low delay is an explicit requirement for applications such as cloud gaming and video conferencing. Delay-based congestion control can achieve the same throughput but significantly smaller delay than loss-based one and is thus ideal for these applications. However, when a delay- and a loss-based flow compete for a bottleneck, the loss-based one can monopolize all the bandwidth and starve the delay-based one. Fair queuing at the bottleneck link solves this problem by assigning an equal share of the available bandwidth to each flow. However, so far no end host based algorithm to detect fair queuing exists. Our contribution is the development of an algorithm that detects fair queuing at flow startup and chooses delay-based congestion control if there is fair queuing. Otherwise, loss-based congestion control can be used as a backup option. Results show that our algorithm reliably detects fair queuing and can achieve low delay and high throughput in case fair queuing is detected. 
\end{abstract}


\maketitle

\glsresetall
\newacronym{cc}{CC}{Congestion Control}
\newacronym{cca}{CCA}{Congestion Control Algorithm}
\newacronym{aqm}{AQM}{Active Queue Management}
\newacronym{rtt}{RTT}{Round Trip Time}
\newacronym{fq}{FQ}{Fair Queuing}
\newacronym{bdp}{BDP}{Bandwidth Delay Product}
\newacronym{qoe}{QoE}{Quality of Experience}

\def\anonymize{no}
\def\tempYes{yes}
\def\tempNo{no}

\section{Introduction}
\label{sec:introduction_fq}

Emerging applications such as cloud gaming \cite{jarschel_evaluation_2011} and remote virtual reality \cite{elbamby_toward_2018} applications require high throughput and low delay. Furthermore, ultra low latency communications have been made a priority for 5G \cite{li_5g_2018}. We argue that for achieving high throughput as well as low delay, congestion control must be taken into account. 

\textit{Delay-based} \glspl{cca} were proposed to provide high throughput and low delay for network flows. They use the queuing delay as an indication of congestion and lower their throughput as soon as the delay increases. Such approaches have been proposed in the last century \cite{brakmo_tcp_1995} and recently have seen a surge in popularity \cite{arun_copa_2018,hock_tcp_2017,mittal_timely_2015,cardwell_bbr:_2016}. While delay-based \glspl{cca} fulfill their goal of low delay and high throughput, they cannot handle competing flows with different \glspl{cca} well \cite{turkovic_fifty_2019, turkovic_interactions_2019}. This is especially prevalent when they compete against loss-based \glspl{cca}, the latter being more ``aggressive''. While the delay-based \glspl{cca} back off as soon as queuing delay increases, the loss-based ones continue increasing their throughput until the buffer is full and packet loss occurs. This results in the loss-based flows monopolize the available bandwidth and starve the delay-based ones \cite{hock_toward_2016,yuan-cheng_lai_improving_2001,awdeh_comparing_2004}.

This unfairness can be mitigated by using \gls{fq} at the bottleneck link, isolating each flow from all other flows and assigning each flow an equal share of bandwidth \cite{dumazet_pkt_sched:_2013}. Consequently, loss-based flows can no longer acquire bandwidth that delay-based flows have released due to their network-friendly behavior. In this case, delay-based \glspl{cca} achieve their goal of high bandwidth and low delay. 

While \gls{fq} solves many problems regarding \gls{cc}, it is still not ubiquitously deployed. Flows can benefit from knowledge of \gls{fq} enabled bottleneck links on the path to dynamically adapt their \gls{cc}. In the case of \gls{fq} they can use a delay-based \gls{cca} and otherwise revert to a loss-based one. Our contribution is the design and evaluation of such a mechanism that determines the presence of \gls{fq} during the startup phase of a flow's \gls{cca} and sets the \gls{cca} accordingly at the end of the startup. 

We show that this solution reliably determines the presence of fair queuing and that by using this approach, queuing delay can be considerably lowered while maintaining throughput high if \gls{fq} is detected. In case no \gls{fq} is enabled, our algorithm detects this as well and a loss-based \gls{cc} is used, which does not starve when facing competition from other loss-based flows. 
While our mechanism is beneficial for network flows in general, we argue that it is most useful for long-running delay-sensitive flows, such as cloud gaming, video conferencing etc. 

To make our work reproducible and to encourage further research, we make the code, the results and the figures publicly available\footnote{\ifx\anonymize\tempYes $\langle$anonymized$\rangle$\else\url{https://github.com/CN-TU/PCC-Uspace}\fi}. 

\section{Related work}
Our work depends on active measurements for the startup phase of \gls{cc} and proposes a new measurement technique to detect \gls{aqm}. This section discusses preliminary work related to these fields. 

\subsection{\gls{cc} using Active Measurements}
Several approaches have been recently proposed which aim to incorporate active measurements into \glspl{cca}. \citep{dong_pcc_2015,dong_pcc_2018} introduced the \gls{cca} \textit{PCC}, which uses active measurements to find the sending rate which optimizes a given reward function. \citep{cardwell_bbr:_2016} introduce the \textit{BBR} \gls{cca}, which uses active measurement to create a model of the network connection. This model is then used to determine the optimal sending rate. An approach proposed by \citep{goyal_elasticity_2018} aims to differentiate bulk transfer flows, which they deem ``elastic'' because they try to grab as much bandwidth as available, from other flows that send at a constant rate, which they call ``unelastic''. They use this elasticity detection to determine if delay-based \gls{cc} can be used or not. If cross traffic is deemed elastic then they use classic loss-based \gls{cc} and otherwise delay-based \gls{cc}. 

\subsection{Flow startup techniques} Besides the classical TCP slow start algorithm \citep{stevens_tcp_1997} some other proposals were made such as Quick Start \citep{jain_quick-start_2019}, which aims to speed up flow start by routers informing the end point, which rate they deem appropriate. \citep{jarvinen_congestion_2019} investigate the impact of \gls{aqm} and \gls{cc} during flow startup. They inspect a variety of different \gls{aqm} algorithms and also investigate \gls{fq}. \citep{kuhlewind_chirping_2010} propose ``chirping'' for flow startup to estimate the available bandwidth. This works by sending packets with different gaps between them to verify what the actual link speed is. \citep{mittal_recursively_2014} propose speeding up flow startup by using packets of different priority. The link is flooded with low-priority packets and in case the link is not fully utilized during startup, these low-priority packets are going to be transmitted, allowing the starting flow to achieve high throughput. If the link is already saturated, the low-priority packets are going to be discarded and no other flow suffers. 

\subsection{Detecting \gls{aqm} mechanisms} There are few publications that investigate methods to detect the presence of an \gls{aqm}. \citep{kargar_bideh_tada_2016} propose a method to detect and distinguish certain popular \gls{aqm} mechanisms. However, unlike our work, it doesn't consider \gls{fq}. \citep{baykal_detection_2017} propose a machine learning based approach to fingerprint \gls{aqm} algorithm. \gls{fq} is not considered, too. 

\subsection{Detecting the presence of a shared bottleneck} Several approaches have been proposed to detect whether in a set of flows, some of them share a common bottleneck \cite{hayes_practical_2014,ferlin_revisiting_2016,hayes_online_2020}. These approaches typically operate by comparing some statistics of flows to check whether they correlate between flows. For example, it can be checked if throughput correlates between two flows. If throughput correlates, it can mean that these flows share a common bottleneck link, and if additional flows join at the bottleneck link, all other flows' throughputs go down, which is the reason why they are correlated. While approaches to detect a shared bottleneck link appear to be similar to what this paper is aiming to do, there are differences: Two flows can share a common bottleneck link but the bottleneck link can have fair queuing. Thus the aims of shared bottleneck and fair queuing detection are different. 

\section{Concept}
\label{sec:concept_fq}

The overall concept is the following:
\begin{enumerate}
\item During flow startup, determine whether the bottleneck link deploys \gls{fq} or not.
\item If \gls{fq} is used change to a delay-based \gls{cca} or revert to a loss-based \gls{cca} otherwise.
\end{enumerate}

\subsection{Algorithm to determine the presence of \gls{fq}}

\begin{figure}
\centering
\subfloat[Sending rate]{\includegraphics[width=0.33\columnwidth]{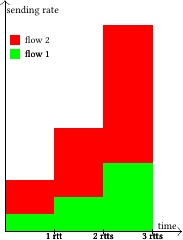}
\label{fig:throughput}}
\subfloat[Receiving rate (no \gls{fq}).]{\includegraphics[width=0.33\columnwidth]{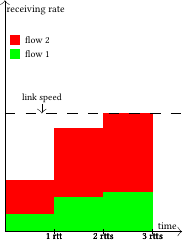}
\label{fig:goodput_no_fq}}
\subfloat[Receiving rate (\gls{fq}).]{\includegraphics[width=0.33\columnwidth]{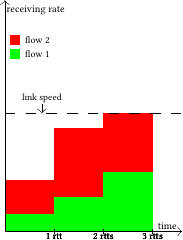}
\label{fig:goodput_fq}}
\caption{An example illustrating our proposed flow startup mechanism. Figure \ref{fig:throughput} shows the sending rate, \ref{fig:goodput_no_fq} the receiving rate in case there's no \gls{fq} and \ref{fig:goodput_fq} the receiving rate if there is \gls{fq}.}
\label{fig:illustration}
\end{figure}

The algorithm to determine the presence of \gls{fq} works as follows (seen from the sender's point of view):
\begin{enumerate}
\item Launch two concurrent flows, flow 1 and flow 2 from the same source host to the same destination host. Initialize flow 1 with some starting sending rate and flow 2 with two times that rate. 
\item After every \gls{rtt}, if no packet loss occurred, double the sending rate of both flows. 
\item If packet loss occurred in both flows in the previous \gls{rtt}, calculate the following metric: 
\begin{align}
\textit{loss ratio} &= \frac{\left(\frac{\textit{receiving rate of flow 1}}{\textit{sending rate of flow 1}}\right)}{\left(\frac{\textit{receiving rate of flow 2}}{\textit{sending rate of flow 2}}\right)}\end{align}

This loss ratio metric indicates if flow 2, which has a higher sending rate, achieves a higher receiving rate (goodput) than flow 1. If this is the case, there's no \gls{fq}, otherwise there is. 

It is necessary to wait for packet loss in both connections because only then it is certain that the link is saturated. Our algorithm assumes that the link is saturated. 
\item If \gls{fq} is used, the $\textit{loss ratio}$ is 2, otherwise it is 1. We choose 1.5 as the cutoff value. 
\item Since the presence or absence of \gls{fq} is now known the appropriate \gls{cca} can be launched. 
\end{enumerate}

Figure~\ref{fig:throughput}, \ref{fig:goodput_no_fq} and \ref{fig:goodput_fq} schematically illustrate this mechanism. We assume a sender being connected to a receiver, with a bottleneck link between them. We also assume that all flows between the source and the destination follow the same route. On this bottleneck link, the queue is either managed by \gls{fq} or by a shared queue (no \gls{fq}). 

Figure~\ref{fig:throughput} shows how the sending rate (observed before the bottleneck link) is doubled after each \gls{rtt}. Furthermore, it shows that flow 2 has double the sending rate of flow 1. 

Figure~\ref{fig:goodput_no_fq} shows the receiving rate at the receiver after the bottleneck link, if this bottleneck link has a a shared queue without \gls{fq}. In the third \gls{rtt} the sending rate exceeds the link speed of the bottleneck and thus the receiving rate is smaller than the sending rate. Because no \gls{fq} is employed, flow 2 manages to have a receiving rate two times the one of flow 1. 

Figure~\ref{fig:goodput_fq} shows the receiving rate of the two flows measured after the bottleneck link in case \gls{fq} is deployed. In the third \gls{rtt}, flow 1 and flow 2 achieve the same receiving rate because of the \gls{fq}, which makes sure that each flow gets an equal share of the total capacity. 

\subsection{Simple delay-based \gls{cc}}

If \gls{fq} is detected, a simple delay-based \gls{cc} is used from then on. Our algorithm is based on the two following simple rules:
\begin{enumerate}
\item If \textit{current rtt} $\leq$ \textit{smallest rtt ever measured on the connection} $+$ 5\textit{ms} then increase the sending rate by 1\%. We choose 5\,ms since a delay of 5\,ms is barely noticeable by humans. This design choice was first established in \citep{nichols_controlling_2012}.
\item Otherwise (\textit{current rtt} $>$ \textit{smallest rtt ever measured on the connection} + 5\textit{ms}) decrease the sending rate by 5\%.
\end{enumerate}

This \gls{cca} is meant as a simple proof-of-concept to demonstrate the efficacy of our \gls{fq} detection mechanism but we do not claim it to be optimal in any way. 

\subsection{Fallback loss-based \gls{cc}}

If the absence of \gls{fq} is detected, a loss based \gls{cc} is used as a fallback. We use the algorithm \textit{PCC} proposed by \citep{dong_pcc_2015}. In simplified terms, this \gls{cca} aims to maximize a utility function, which strives for high throughput and low packet loss. However, it is reasonably aggressive and does not starve when competing with other loss-based \glspl{cca} such as Cubic \citep{ha_cubic_2008}.

\section{Implementation}

We base our implementation on the code of \textit{PCC Vivace} \citep{dong_pcc_2018}, which in turn is an improved version of PCC. The PCC code is based on the \textit{UDT} library \citep{gu_udt_2007}, which is a library that provides a reliable transport protocol similar to TCP on top of UDP. This approach is similar to QUIC \citep{iyengar_quic_2018}. An advantage of implementing our approach on top of UDP as opposed to TCP is that it is easier to develop and modify since no kernel module needs to be created and that it is more secure since the code is isolated in a user space process and most attacks can only influence this process and not the entire kernel. 

\subsection{Deployment}

Our approach of detecting \gls{fq} relies on using two concurrent flows between the sender and the receiver. During flow startup, user data is transmitted in two flows that the receiver must detect and reassemble. This may be a challenge for deployments since solutions that require cooperation of several instances are more difficult to deploy at a wide scale in the Internet. 
However, multipath TCP \citep{handley_tcp_2012} and QUIC, for which there are proposals to enable recombining data of several flows \citep{de_coninck_multipath_2017}, are becoming increasingly prevalent. We thus argue that the actual obstacles to deployment of our solution are constantly decreasing since QUIC (mostly Google traffic) already accounts for around 
$10\%$ of all Internet traffic \citep{ruth_first_2018} and is growing. Moreover, Apple's macOS and iOS support multipath TCP \citep{apple_about_2018} support. 

\section{Experiment setup}

For the evaluation we use the network emulation library \texttt{py-virtnet}\footnote{\url{https://pypi.org/project/py-virtnet/}}. All experiments run on virtual hosts on one physical host, on which the necessary network components and virtual hosts are emulated using Linux network namespaces. The setup consists of two virtual hosts connected through a virtual switch. One virtual host acts as a server and one as a client and a bulk transfer is initiated from the client to the server. 

For the experiments with \gls{fq} we use the \texttt{fq} module for Linux. This module keeps one drop tail queue for each flow with a configurable maximum size of packets. However, the design of our \gls{fq} detection algorithm does not assume any particular queue management algorithm being used for the queue of each flow and thus is also compatible with more sophisticated \gls{aqm} mechanisms which were explicitly developed to be used in conjunction with \gls{fq} \citep{taht_flow_2018,hoiland-jorgensen_piece_2018,bachl_cocoa_2019,bachl_lfq_2020}. It only assumes that each flow gets the same share of bandwidth and that for a flow, whose queue is growing faster because more packets of this flow arrive at the bottleneck, packets are dropped faster than for a flow of which fewer packets appear at the bottleneck link. For the experiments with a shared queue we use \texttt{pfifo}, which is a simple drop tail queue which can contain up to a fixed number of packets. 

\section{Results}

\subsection{Delay- vs.~loss-based \gls{cc}}

\begin{figure}[h]
\centering
\subfloat[Throughput (receiving rate)]{%
  \includegraphics[width=0.48\columnwidth]{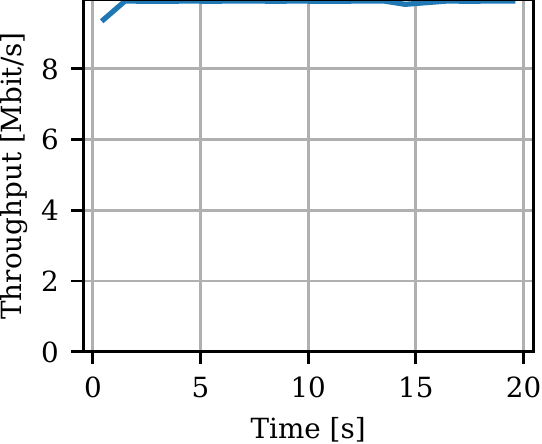}%
  \label{fig:just_one_flow_tcp_throughput}
}\hfill
\subfloat[\gls{rtt}]{%
  \includegraphics[width=0.48\columnwidth]{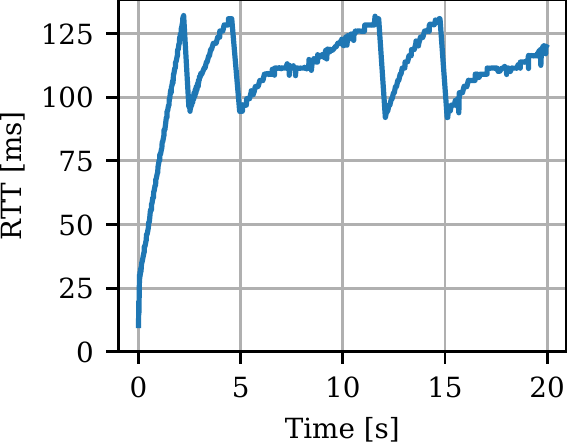}%
  \label{fig:just_one_flow_tcp_delay}
}
\caption{Throughput and delay of a flow controlled by the loss-based \gls{cca} Cubic on a link with a speed of 10\,Mbps, a delay of 10\,ms and a buffer size of 100 packets (\texttt{pfifo}'s default value).}
\label{fig:just_one_flow_tcp}
\end{figure}

\begin{figure}[h]
\centering
\subfloat[Throughput (receiving rate)]{%
  \includegraphics[width=0.48\columnwidth]{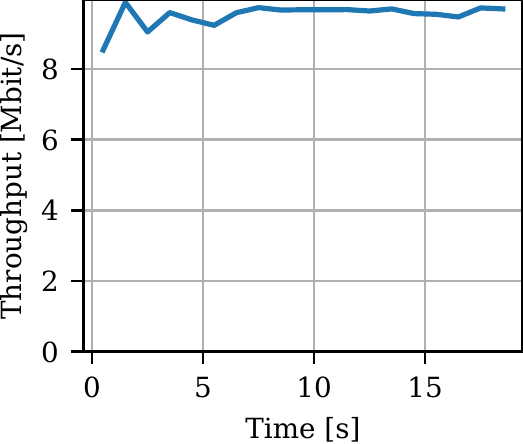}%
}\hfill
\subfloat[\gls{rtt}]{%
  \includegraphics[width=0.48\columnwidth]{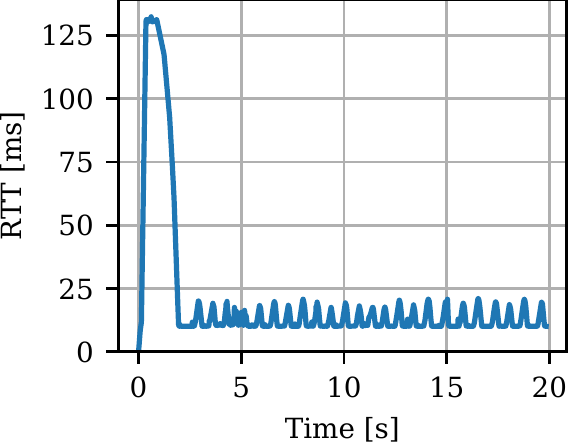}%
}
\caption{Throughput and delay of a flow controlled by our delay-based \gls{cca} on a link with a speed of 10\,Mbps, a delay of 10\,ms and a buffer size of 100 packets.}
\label{fig:just_one_flow_udp}
\end{figure}

First, we demonstrate empirically that delay-based \gls{cc} actually has a benefit over loss-based one. On one side, \autoref{fig:just_one_flow_tcp} shows that the loss-based \gls{cc} achieves a throughput of almost 10\,Mbps on a 10\,Mbps link, which means full utilization and an \gls{rtt} that is  higher than 100\,ms, which is not desirable. On the other side, \autoref{fig:just_one_flow_udp} shows an example flow using our simple delay-based \gls{cca} and a slow start algorithm similar to the one used by the classic TCP \gls{cca} \textit{New Reno} \citep{allman_tcp_1999}. While the delay-based flow also achieves very high utilization, delay rarely exceeds 20\,ms, meaning that it is more than 5 times lower than the one of the loss-based \gls{cca} of \autoref{fig:just_one_flow_tcp}. The initial peak in \autoref{fig:just_one_flow_udp} is caused by the exponential increase of the sending rate in the beginning, which is commonly used by transport protocols and called ``slow start'' \citep{stevens_tcp_1997}. Our delay-based \gls{cc} also uses a mechanism similar to slow start in the beginning. 

\subsection{Comparing flows with different Congestion Controls and AQM mechanisms}
In this scenario, we show an example of a flow with our mechanism that shares a bottleneck link with a flow using a loss-based congestion control. 

\subsubsection{No \gls{fq}}
\begin{figure}[h]
\centering
\subfloat[Throughput (receiving rate)]{%
  \includegraphics[width=0.48\columnwidth]{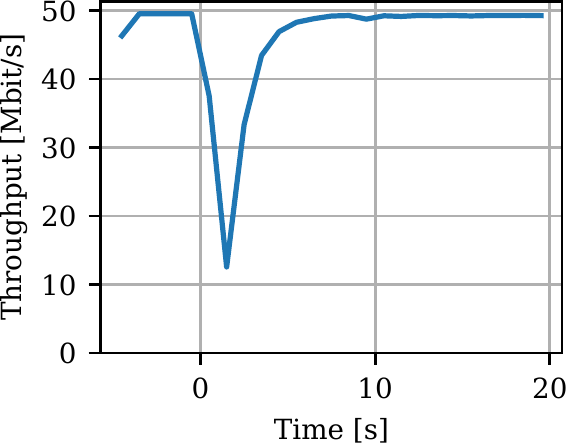}%
}\hfill
\subfloat[\gls{rtt}]{%
  \includegraphics[width=0.48\columnwidth]{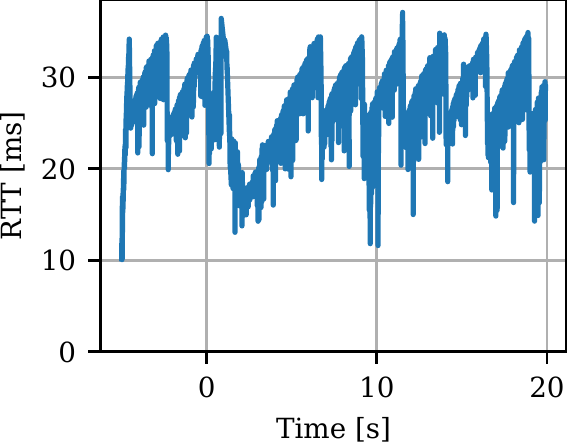}%
}
\caption{Throughput and delay of a flow controlled by the loss-based \gls{cca} Cubic on a link with a speed of 50\,Mbps, a delay of 10\,ms and a buffer size of 100 packets. The bottleneck is controlled by a shared queue (no \gls{fq}) and is shared with the flow of \autoref{fig:pfifo_udp}.}
\label{fig:pfifo_tcp}
\end{figure}

\begin{figure}[h]
\centering
\subfloat[Throughput (receiving rate)]{%
  \includegraphics[width=0.48\columnwidth]{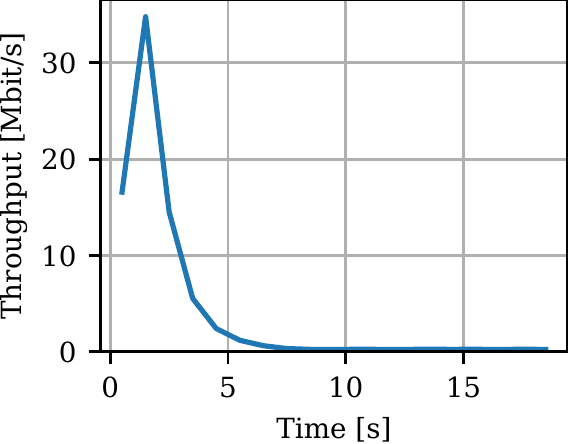}%
}\hfill
\subfloat[\gls{rtt}]{%
  \includegraphics[width=0.48\columnwidth]{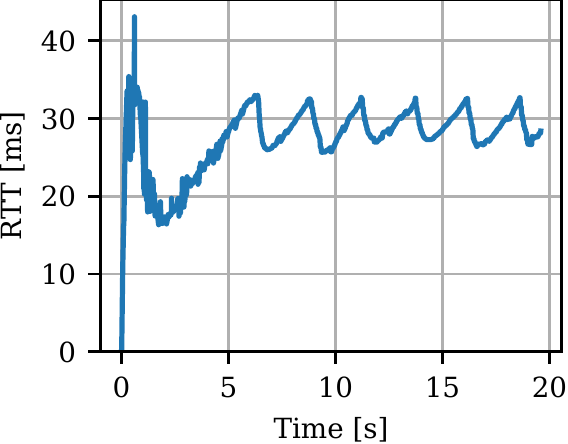}%
}
\caption{Throughput and delay of a flow controlled by our delay-based \gls{cca} on a link with a speed of 50\,Mbps, a delay of 10\,ms and a buffer size of 100 packets. The bottleneck is controlled by a shared queue (no \gls{fq}) and is shared with the flow of \autoref{fig:pfifo_tcp}.}
\label{fig:pfifo_udp}
\end{figure}

The examples in \autoref{fig:pfifo_tcp} and in \autoref{fig:pfifo_udp} show the loss-based \gls{cca} Cubic compete against our delay-based \gls{cca} on a link without \gls{fq}. It is important to note that we do not use our \gls{fq} detection mechanism here but instead choose delay-based \gls{cc} from the beginning to demonstrate the starvation of delay-based flows that happens when they compete against loss-based ones without \gls{fq}. We start the loss-based flow before to make sure that it has sufficient time to fill up the link: The loss-based flow starts 5 seconds earlier and after 5 seconds, the delay-based flow joins. We choose these 5 seconds so that the first flow can enter its stable state because we want to show how how the flow, even though it had the advantage of the 5 seconds in which it could use the link all for itself, loses to the loss-based \gls{cc}: While the delay-based flow gains some share of the link during startup, it is pushed away by the loss-based flow later on and ``starves''. 

\subsubsection{\gls{fq}}

\begin{figure}[h]
\centering
\subfloat[Throughput (receiving rate)]{%
  \includegraphics[width=0.48\columnwidth]{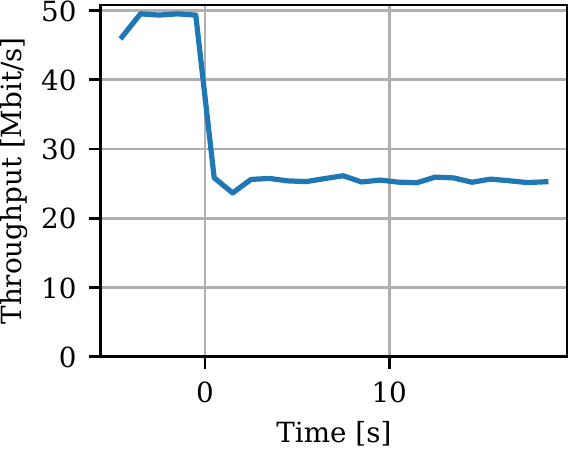}%
}\hfill
\subfloat[\gls{rtt}]{%
  \includegraphics[width=0.48\columnwidth]{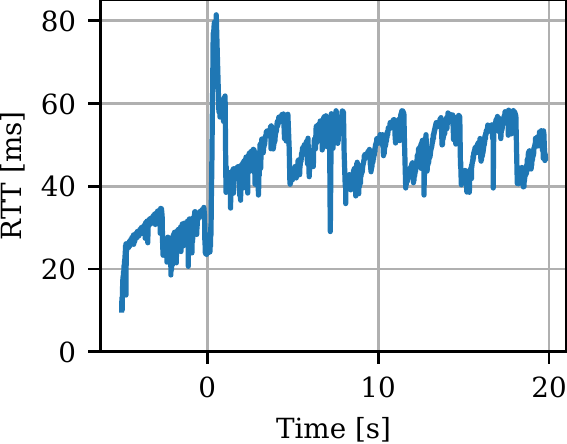}%
}
\caption{Throughput and delay of a flow controlled by the loss-based \gls{cca} Cubic on a link with a speed of 50\,Mbps, a delay of 10\,ms and a buffer size of 100 packets. The bottleneck is controlled by \gls{fq} and is shared with the flow of \autoref{fig:fq_udp}.}
\label{fig:fq_tcp}
\end{figure}

\begin{figure}[h]
\centering
\subfloat[Throughput (receiving rate)]{%
  \includegraphics[width=0.48\columnwidth]{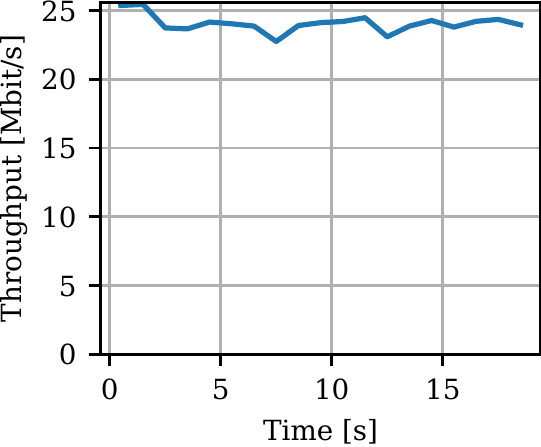}%
}\hfill
\subfloat[\gls{rtt}]{%
  \includegraphics[width=0.48\columnwidth]{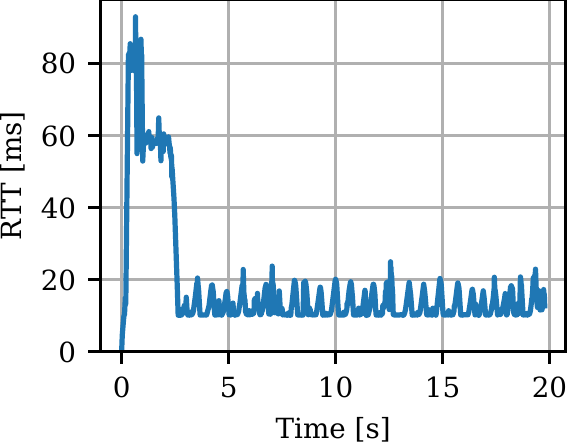}%
}
\caption{Throughput and delay of a flow controlled by our delay-based \gls{cca} on a link with a speed of 50\,Mbps, a delay of 10\,ms and a buffer size of 100 packets. The bottleneck is controlled by \gls{fq} and is shared with the flow of \autoref{fig:fq_tcp}.}
\label{fig:fq_udp}
\end{figure}

For a link with fair queuing, the following example results show that throughput is identical for our flow (\autoref{fig:fq_udp}) and the Cubic flow (\autoref{fig:fq_tcp}), just that the Cubic flow's delay is more than two times higher than the delay-based one's. Thus, our mechanism detects that the bottleneck uses \gls{fq} and uses a delay-based congestion control because it is better suited in a scenario with \gls{fq}. 

\subsection{Accuracy of the \gls{fq} detection mechanism}

To evaluate if our proposed mechanism correctly recognizes \gls{fq} in a systematic manner, we perform experiments using a wide range of network configurations: we vary bandwidth from 5 to 50\,Mbps, delay from 10 to 100\,ms and the buffer size from 1 to 100 packets, using a grid of these parameters. For each parameter, we take 5 values, evenly spaced. Thus, we run experiments with 125 different configurations. We run each configuration once with \gls{fq} and once without. Results show that for experiments without \gls{fq} (using a shared queue), in over 99\% of cases, the absence of \gls{fq} was correctly detected and in less than 1\% of cases, our algorithm detected \gls{fq} even though there was no \gls{fq}. For experiments, during which \gls{fq} was actually deployed on the bottleneck link, this was correctly detected in 97\% of cases and wrongly in 3\% of the time. The overall accuracy is thus 98\%, which we consider sufficient. 

To verify the behavior of our algorithm under the presence of cross traffic, we repeated the above experiments but with one bulk traffic flow using Cubic running at the same time, which we start 5 seconds before so that it has sufficient time to gain full bandwidth. Also under these circumstances we also achieve 98\% detection accuracy using our algorithm. We investigated the reason for the detection accuracy not being 100\% and found the following: The PCC code relies on packet scheduling using timers, which is very computationally expensive because it needs a timer for every packet and thus the program has to be put to sleep/woken up repeatedly by the operating system. Thus, if processor load is high, the timers fail and PCC can't send as many packets as it would want to. our algorithm relies on flow 1 sending half the rate of flow 2 but this cannot always be guaranteed if CPU load is high. We tried to minimize all tasks which run in the background and achieved the accuracy we got. When we ran the experiments while we was browsing the web, we got lower accuracy. Solutions to the problem are to not schedule packets using timers or to implement the algorithm not in user space but in kernel space. Then the computational cost of could decrease because fewer context switches between the PCC user space application and the kernel are necessary. 

\subsection{Systematic evaluation}
 
Besides the examples we provided, we also conducted a more systematic comparison of the throughput and delay that our algorithm achieves: we use the same scenario as above and again vary bandwidth from 5 to 50\,Mbps, delay from 10 to 100\,ms and the buffer size from 1 to 100 packets. For each parameter, we take 5 values, evenly spaced, totaling 125 distinct scenarios. We use \gls{fq} at the bottleneck link. We run each scenario both using Cubic as well as our algorithm for 30 seconds each. We average the achieved throughputs and average delays from all experiments for both Cubic as well as our algorithm. For Cubic, the mean throughput is 21\,Mbit/s and the mean \gls{rtt} is 96\,ms while for our algorithm it is 26\,Mbit/s and 64\,ms. This shows that our algorithm can correctly determine that there is \gls{fq} and switch to the delay-based \gls{cc} and that our algorithm achieves significantly lower delay (loss-based Cubic's delay is 50\% higher) and -- surprisingly -- higher throughput. This is astonishing considering that loss-based \glspl{cca} such as Cubic should be more aggressive in filling the link. When analyzing the results we identified that Cubic performs very poorly in scenarios with small buffers because the multiplicative decrease of Cubic is 0.7, meaning that the bandwidth is reduced by 30\% after every packet loss while our algorithm only reduces it by only 5\%. Thus Cubic reduces bandwidth too much after packet loss which results in underutilization of the link. 

\section{Discussion}

During the implementation of our algorithm we noticed peculiar behavior when not using \gls{fq}: During startup, we make flow 1 have half the sending rate of flow 2. We would expect the receiving rates to be the same: flow 1 achieving around half the rate of flow 2. However, we noticed that usually flow 1 would have a receiving rate of 0. After investigating the issue we noticed that the reason was pacing: PCC (and other \glspl{cca}) such as BBR send packets in regular intervals, for example every millisecond, because sending regularly keeps the queues shorter. This behavior is called pacing. However, when there are two flows where one flow's sending rate is a multiple of the other's, an interleaving effect occurs: flow 1 would always send each packet slightly after flow 2 sends a packet. Then, when the packets arrive at the bottleneck, flow 2's packet, which arrives a bit earlier, fills up the buffer and flow 1's packet is consequently always dropped. As a solution, we do not send packets at regular intervals during startup but always add a small random value: For example, if the sending interval were 1\,ms, we would send the packet randomly in the interval from 0.5 to 1.5\,ms. This solves the aforementioned problem. The insight we gained from this is that while pacing is generally supportive, it is actually harmful if several flows compete at a bottleneck without \gls{fq} and one flow's sending rate is a multiple of the other's.

One problem of our proposed solution occurs when the bottleneck link changes while a flow is running. For example, the bottleneck link could be the home router in the beginning, while later the bottleneck link could be another router deeper in the core network. The problem occurs if the home router supports \gls{fq} and our algorithm detects this while the router in the core network, which becomes the bottleneck later on, doesn't support \gls{fq}. Then, our algorithm is going to use a delay-based \gls{cc} because that's what it determined to be correct in the beginning of the flow. However, later when the bottleneck changes, our algorithm is still using delay-based \gls{cc} even though the current bottleneck doesn't support \gls{fq} anymore. As a result, it might be possible that our algorithm uses the wrong \gls{cc}. One way to prevent this problem from occurring is to use periodic measurements. For example, our algorithm to determine \gls{fq} could be used every 5 or 10 seconds to measure again if the current bottleneck supports \gls{fq}. 

Another scenario that is worth noting is how our proposed solution deals with load balancing: In this research we assume that both of our flows, which we use during startup to determine the presence of \gls{fq} take the same path to the host on the other side. However, with load balancing it could happen that flow 1 takes a different path than flow 2. Fortunately, several load balancing algorithms have been proposed whose aim is to make sure that connections, which use multiple paths using Multipath TCP or QUIC, stay on the same path \citep{olivier_bonaventure_multipath_2018,olteanu_datacenter_2016,lienardy_towards_2016}. Since we envision our solution to be based on Multipath TCP or QUIC load balancing should not be a problem. \textit{Apple Music}, for example, successfully load balances multipath connections \citep{olivier_bonaventure_apple_2019}. 

An alternative to our algorithm could be that all routers inform the endpoints if they support \gls{fq} or not. Such a solution could, for example, use an IP or a TCP extension, which every router uses to indicate if they're capable of \gls{fq}. Then, if every device on a path supports \gls{fq}, the client would use a delay-based \gls{cca}. However, while this solutions would achieve good results, the problem is that solutions, which require participation of every device in the Internet generally suffer from insufficient deployment because not every hardware vendor and network operator can be forced to implement the proposed solution for informing about \gls{fq} capability. 

Concluding, we argue that our new flow startup method can help to increase the deployment of delay-based \gls{cc}. Especially for applications that are very delay-sensitive, such as video conferencing and cloud gaming, the adoption of delay-based \gls{cc} can result in a significant increase of \gls{qoe}. We demonstrated that our method works using a prototype implementation using PCC but we are also confident that a similar flow startup method can also be integrated with other \glspl{cca} such as BBR. 

\renewcommand*{\bibfont}{\small}
\bibliographystyle{ieeetr}
\bibliography{fq_detection}

\end{document}